\documentstyle[preprint,aps,psfig]{revtex}
\newcommand{\beq}{\begin{equation}}
\newcommand{\eeq}{\end{equation}} 
\newcommand{\beqa}{\begin{eqnarray}}
\newcommand{\eeqa}{\end{eqnarray}}
\newcommand{\ba}{\begin{array}}
\newcommand{\ea}{\end{array}}

\begin{document}
\draft
\widetext 
\title{Triaxial Bright Solitons in Bose-Condensed Atomic Vapors} 
\author{Luca Salasnich} 
\address{Istituto Nazionale per la Fisica della Materia, 
Unit\`a di Milano, \\ Dipartimento di Fisica, Universit\`a di Milano, \\
Via Celoria 16, 20133 Milano, Italy \\
E-Mail: salasnich@mi.infm.it} 
\maketitle

\begin{abstract} 
The properties of triaxial bright solitons (TBSs) 
made of attractive Bose-Einstein condensed atoms under transverse 
anisotropic harmonic confinement are investigated 
by using a variational approach. 
We show that these metastable TBSs change their shape 
from cigar-like to disc-like by increasing the inter-atomic strength. 
Moreover, we find that the collective oscillations 
of a TBS strongly depend on the anisotropy parameter of the 
external potential. We calculate in detail 
the properties of TBS close to the collapse, which sets up 
at a critical value of the inter-atomic strength. 
This critical strength is maximal in the isotropic 
(axially symmetric) case and slightly reduces by increasing 
the anisotropy. Finally, we investigate the formation 
of multiple TBSs via modulational instability induced 
by a sudden change of the scattering length from positive to negative. 
\end{abstract}


\narrowtext 

\newpage 

Bright solitons (BSs) have been recently  
obtained in two experiments with Bose-Einstein condensates (BECs) 
of $^7$Li dilute vapors at ultra-low temperatures [1,2]. 
In these experiments the attractive BEC 
is confined in two 
directions by a transverse cylindric isotropic 
harmonic potential and travels in the 
third direction, the longitudinal cylindric axis, 
without a relevant spreading. The theoretical analysis of these 
BS configurations has been developed by 
various authors [3-8]. It has been shown that the confinement 
in two directions is necessary to create single [3-6] 
or multiple [7,8] metastable BSs, which collapse 
above a critical number of particles [5,6,8]. 
In the previous theoretical investigations only an isotropic 
transverse cylindric harmonic confinement has been studied. 
Under these conditions the BS 
is cigar-shaped, it becomes less cigar-shaped by increasing 
the inter-atomic strength and it acquires a quasi-spherical shape 
only very close to the critical inter-atomic strength of 
the collapse [5,6]. 
\par 
In this paper we consider the case of an 
attractive BEC under {\it anisotropic} 
harmonic confinement. 
This condensate BS is triaxial and we find that 
it can be cigar-shaped or disc-shaped by changing 
the inter-atomic strength. 
The triaxial bright soliton (TBS) collapses 
at a critical inter-atomic strength 
that decreases by increasing the anisotropy parameter of the 
transverse harmonic potential. 
By using a variational approach we calculate the collective 
oscillations of the TBS. We find that close to the collapse 
the longitudinal collective frequency reaches its maximum value 
and then quickly goes to zero. The maximum value 
of the longitudinal collective frequency strongly depends on 
the anisotropy parameter. We calculate also the number 
of TBSs which can be obtained, starting with a repulsive and 
axially uniform BEC, by suddenly changing the 
scattering length from positive to negative. 
This mechanism, known as modulational instability [7,8], 
has been experimentally applied to get a soliton train 
of axially symmetric BSs [2]. 
\par 
Nowadays BECs are routinely created 
and trapped in magnetic or 
optical traps. Such traps can be modeled with good accuracy 
with harmonic potentials. Here we study an attractive BEC 
with $N$ atoms confined in the cylindric transverse direction 
by an anisotropic harmonic potential 
$U({\bf r}) = (m/2)(\omega_1^2 x^2 +\omega_2^2 y^2 )$,  
where $m$ is the mass of a Bose-condensed atom and 
$\omega_1$ and $\omega_2$ are the two frequencies of the 
transverse confinement. The ratio $\lambda = \omega_2/\omega_1$ 
is the anisotropy parameter of the harmonic potential. Note that 
there is no confinement along the z axis. 
\par 
The scaled Lagrangian density of this attractive 
BEC is given by 
\beq 
{\cal L} = i \psi^* \partial_t \psi + 
{1\over 2}\psi^* \nabla^2\psi - 
U |\psi|^2 - 2\pi {a_s\over a_H} |\psi|^4  \; , 
\eeq 
where $\psi({\bf r},t)$ is the macroscopic wave function of 
the BEC normalized to $N$, $a_s$ is the s-wave scattering 
length ($a_s<0$ with attractive atoms) and 
$a_H =(\hbar/(m\omega_H))^{1/2}$ is the average harmonic length 
of the trapping potential with $\omega_H=(\omega_1\omega_2)^{1/2}$. 
Note that in Eq. (1) lengths are in units of $a_{H}$, 
energies are in units of $\hbar \omega_H$, 
and the time is in units of $\omega_H^{-1}$. 
In the rest of the paper we use these scaled variables. 
\par 
The Euler-Lagrange equation obtained from 
the Lagrangian density $\cal L$ is the familiar 3D 
time-dependent Gross-Pitaevskii equation (GPE), which describes 
very accurately a pure BEC of Bosonic vapors [9]. 
By using the 3D GPE one finds that 
the ground-state of an attractive BEC 
is a collapsed state of zero radius and infinite negative energy but, 
as previously stressed, metastable states can be obtained, 
up to a critical number of particles, under external confinement 
in two or three directions [5,6]. The stationary metastable states 
of an attractive BEC under confinement in two directions are 
referred as BSs because they are self-confined in the third 
direction and can propagate without spreading along that 
direction [10]. For $\lambda =1$ one has axial BSs, whose 
statical and dynamical properties have been analyzed in detail 
in previous papers [5-8]. For $\lambda \neq 1$ one gets TBSs, 
whose properties are investigated in this paper. 
\par 
Instead of numerically solving the full 3D GPE 
we investigate the properties of the TBS 
by using a variational approach [11], which is 
computationally much less demanding. 
Our variational ansatz for the macroscopic wave function 
of the attractive BEC is the following 
\beq 
\psi({\bf r}, t) 
= \prod_{k=1}^3 {N^{1/6} \over \left( \pi \sigma_k(t)^2 \right)^{1/4} }  
\exp{\left\{  - { x_k^2 \over 2 \sigma_k(t)^2} + i \beta_k(t) x_k^2 
\right\} }   
\eeq 
with ${\bf r}=(x_1,x_2,x_3)=(x,y,z)$. $\sigma_k(t)$ and $\beta_k(t)$ 
are the time-dependent variational parameters. 
The $\sigma_k(t)$ are the widths of the TBS in the three axial 
directions. Note that in order to describe the time evolution 
of the variational wave function, the phase factor 
$i\beta_k(t) x_k^2$ is needed. 
\par 
By using this Gaussian trial wave function, after spatial 
integration of the Lagrangian density $\cal L$ of Eq. (1), 
one finds a new effective Lagrangian $L$ that has 
$\sigma_k(t)$ and $\beta_k(t)$ as generalized coordinates. 
The three Euler-Lagrange equations of $L$ with respect to 
$\sigma_k(t)$ are given by 
\beqa 
{\ddot \sigma}_1 + \lambda_1^2 \sigma_1 = {1\over \sigma_1^3} 
- \sqrt{2\over \pi} 
{\gamma \over \sigma_1^2 \sigma_2 \sigma_3} \; , 
\nonumber 
\\
{\ddot \sigma}_2 + \lambda_2^2 \sigma_2 = {1\over \sigma_2^3} 
- \sqrt{2\over \pi} 
{\gamma \over \sigma_1 \sigma_2^2 \sigma_3} \; , 
\\
{\ddot \sigma}_3 = {1\over \sigma_3^3} 
- \sqrt{2\over \pi} 
{\gamma \over \sigma_1 \sigma_2 \sigma_3^2} \; ,  
\nonumber 
\eeqa 
where $\gamma =N|a_s|/a_H$ is the scaled inter-atomic strength and  
$\lambda_1=\omega_1/\omega_H=1/\sqrt{\lambda}$ and 
$\lambda_2=\omega_2/\omega_H=\sqrt{\lambda}$ 
are the scaled frequencies of the transverse harmonic potential. 
The three Euler-Lagrange equations of $L$ with respect to 
$\beta_k(t)$ give instead the following expressions: 
$\beta_k = - {\dot \sigma}_k /(2\sigma_k)$, with $k=1,2,3$. 
Thus the time dependence of $\beta_k(t)$ is fully determined 
by that of $\sigma_k(t)$. 
\par 
The stationary metastable states of Eqs. (3), namely the TBS 
static configurations, are found by setting 
${\ddot \sigma}_k = 0$. The resulting algebraic equations can be 
easily solved numerically. The solutions correspond 
to metastable states if they are local minima of the effective 
potential energy 
$$
W(\sigma_1,\sigma_2,\sigma_3) = 
{1\over 2} \left( {1\over \sigma_1^2} + {1\over \sigma_2^2} 
+ {1\over \sigma_3^2} \right) 
$$
\beq 
+ {1\over 2} \left( \lambda_1^2 \sigma_1^2 + 
\lambda_2^2 \sigma_2^2 \right) 
- \sqrt{2\over \pi} {\gamma \over \sigma_1\sigma_2\sigma_3 } 
\eeq 
associated to the Newtonian Eqs. (3). 
In Figure 1 we plot the three widths $\sigma_1$, $\sigma_2$ and 
$\sigma_3$ of the metastable TBS as a function of the inter-atomic 
strength $\gamma$. 
Note that the transverse widths $\sigma_1$ and $\sigma_2$ 
do not change very much with respect to the non-interacting values 
$\sigma_1=1/\sqrt{\lambda_1}=\lambda^{1/4}$ and 
$\sigma_2=1/\sqrt{\lambda_2}=1/\lambda^{1/4}$, 
while the longitudinal width $\sigma_3$ is divergent for 
$\gamma =0$ and approaches one of the two transverse 
widths for $\gamma$ close to the critical value $\gamma_c$. 
For $\gamma > \gamma_c$ there are no more metastable solutions: 
one has the so-called collapse of the attractive BEC. 
\par
It is important to observe that, by using the Gaussian 
variational approach, the critical value $\gamma_c$ of 
the collapse is slightly overestimated with respect to 
the ``exact'' numerical calculations. For instance, 
with $\lambda =1$ the ``exact'' result gives 
$\gamma_c = 0.676$ [12] while our variational method 
predicts $\gamma_c=0.778$. Moreover, in the case of 
asymmetric harmonic trap along the three axes, it has been 
shown that the relative error of $\gamma_c$ by using 
the variational approach is always within $10$\% [12]. 
\par
Figure 1 shows that for small values of $\gamma$ the TBS is 
highly cigar-shaped but near $\gamma_c$ the TBS becomes 
disk-shaped due to the enormous compression along the 
longitudinal z axis. In particular, close to the collapse 
with $\lambda =1$ the system is 
spherical-shaped but choosing a large anisotropy 
($\lambda \gg 1$ or $\lambda \ll 1$) it is strongly disk-shaped. 
Note that, due to the symmetry of the problem, one has 
$\sigma_1(\lambda) = \sigma_2(\lambda^{-1})$ and 
$\sigma_3(\lambda)= \sigma_3(\lambda^{-1})$. 
\par 
The Gaussian approximation of the TBS wave function 
can be used to study also the collective oscillations the TBS. 
The diagonalization of the 
Hessian matrix $\partial^2 W/\partial\sigma_k \partial \sigma_l$ 
of the effective potential energy $W$ of Eq. (4) 
gives three frequencies $\Omega_1$, $\Omega_2$ 
and $\Omega_3$ of collective excitations around the TBS solution. 
The $3\times 3$ Hessian matrix can be numerically diagonalized 
by choosing, for fixed $\gamma$ and $\lambda$, 
the widths $\sigma_k$ that satisfy the Eqs. (3) with 
${\ddot \sigma}_k=0$. 
\par 
In Figure 2 we plot such frequencies. 
Only for $\gamma=0$ the frequencies 
$\Omega_1$, $\Omega_2$ and $\Omega_3$ can be interpreted 
as collective oscillations along $x$ axis, $y$ axis and $z$ axis, 
respectively; nevertheless 
the mixing angle remains quite small also for finite values of $\gamma$ 
so they can be associated to the motion along the three axes. 
The transverse frequencies $\Omega_1$ and $\Omega_2$ are 
practically constant with respect to $\gamma$ and 
equal to the non-interacting values $\Omega_1=2\lambda_1 
= 2/\sqrt{\lambda}$ and $\Omega_2=2\lambda_2=2\sqrt{\lambda}$;  
only very close to $\gamma_c$ the two transverse collective 
frequencies suddenly grows. Instead, 
the longitudinal frequency $\Omega_3$ is zero 
for $\gamma=0$, it increases with $\gamma$ but close 
to $\gamma_c$ it falls down to zero. 
Note that, again due to the symmetry of the problem, 
one has $\Omega_1(\lambda) = \Omega_2(\lambda^{-1})$ 
and $\Omega_3(\lambda)= \Omega_3(\lambda^{-1})$. 
The maximum value $\Omega_3^{(m)}$ of the frequency 
$\Omega_3$ is plotted in Figure 3 (dot-dashed line) 
as a function of $\lambda$. 
In Figure 3 is also shown (full line) 
the graph of $\gamma_c(\lambda)$. 
Both $\Omega_3^{(m)}$ and $\gamma_c$ slowly decreases 
by increasing the anisotropy (note the logarithmic scale 
in the $\lambda$ axis of Figure 3). 
\par 
In a recent experiment [2] a train of bright solitons 
has been obtained with an initially attractive and axially uniform 
BEC by a sudden change of the scattering 
length $a_s$ from positive (repulsive) to negative (attractive),  
induced by an external magnetic field (Feshbach resonance [13]).  
The formation of this multi-soliton configuration can be explained 
as due to the modulational instability 
of the time-dependent wave function of the BEC, 
driven by imaginary Bogoliubov excitations [7,8]. 
An analytical formula for the number $N_s$ of bright solitons 
generated with a quasi-1D condensate 
via modulational instability has been derived by 
Al Khawaja {\it et al.} [7] and by Salasnich, Parola 
and Reatto [8]. The formula has been generalized by Salasnich [8] 
to the case of a 3D BEC under isotropic transverse 
confinement. Here we predict that the anisotropy 
of the transverse harmonic confinement strongly affects 
the number $N_s$ of bright solitons. 
\par 
Let us consider a BEC made of $N$ atoms 
with positive scattering length ($a_s>0$) trapped by a harmonic 
potential in the transverse direction ($x,y$) and 
by a box potential of length $L$ in the longitudinal 
axial direction ($z$). 
In this way the stationary BEC is uniform 
along the z axis (axially uniform) and it can be modeled by 
the following variational wave function
\beq 
\psi({\bf r}) 
= n^{1/2} {1\over \left( \pi \sigma_1 \sigma_2 \right)^{1/2} }  
\exp{\left\{  - \left( {x^2 \over 2 \sigma_1^2} 
+ {y^2 \over 2 \sigma_2^2} \right) \right\} } \; , 
\eeq 
where $n=N/L$ is the axial density of the BEC 
and the two variational parameters $\sigma_1$ and $\sigma_2$ 
are the width of the BEC along $x$ and $y$ axes. 
It is straightforward to derive the energy 
of this BEC from the Lagrangian density of Eq. (1) 
by setting $\partial_t\psi =0$, inserting the Eq. (5) into Eq. (1) 
and integrating the lagrangian density 
over $x$ and $y$. In this way one finds for the energy 
the following formula 
\beq 
E = {g \; n^2\over 2\sigma_1 \sigma_2} + 
{1\over 4} \left( {1\over \sigma_1^2} + {1\over \sigma_2^2} + 
\lambda_1^2 \sigma_1^2 + \lambda_2^2 \sigma_2^2 \right) n \; ,  
\eeq 
where $g=2a_s/a_H$. This energy depends on the two variational 
parameters $\sigma_1$ and $\sigma_2$. 
The minimization of the energy 
$E$ with respect to $\sigma_1$ and $\sigma_2$ 
gives the equations  
\beq 
\lambda_1^2 \sigma_1^4 = 1 + g n {\sigma_1 \over \sigma_2} \; , 
\;\;\;\;\; 
\lambda_2^2 \sigma_2^4 = 1 + g n {\sigma_2 \over \sigma_1} \; . 
\eeq 
From these nontrivial equations one numerically finds, 
for a fixed inter-atomic strength $g$, the transverse widths 
$\sigma_k$ of the Bose gas as a function 
of the axial density $n$. 
Having derived the functions $\sigma_k(n)$, one 
can numerically determine the chemical potential 
$\mu={\partial E/\partial n}$. From the chemical 
potential $\mu$ one gets the sound velocity $c$ in the 
longitudinal axial direction, which satisfies the equation 
$c = (n {\partial \mu /\partial n})^{1/2} 
= ( n {\partial^2 E /\partial n^2} )^{1/2}$. 
It is important to stress that the energy $E$, the chemical 
potential $\mu$ and the sound velocity $c$ obtained with 
Eqs. (6-7) give precisely the 1D GPE results in 
the weak-coupling 1D limit ($\sigma_1=\lambda^{1/4}$ and 
$\sigma_2=1/\lambda^{1/4}$) while they have the correct 
(Thomas-Fermi) 3D GPE behaviour in the strong-coupling 3D limit 
($\sigma_1^4=\lambda^2 gn$ and $\sigma_2^4 = gn/\lambda^2$) 
(see also [14] for the $\lambda=1$ case). 
\par 
The sound velocity is useful to calculate 
the Bogoliubov excitations 
$\epsilon_k =[(k^2/2)\left(k^2/2 + 2 c^2 \right)]^{1/2}$ 
of the axially uniform BEC. 
By suddenly changing the scattering length $a_s$ to a 
negative value, the excitations frequencies corresponding to 
$k<k_c = 2 |c|$ become 
imaginary and, as a result, small perturbations grow 
exponentially in time. The maximum rate of growth is at 
$k_0=k_c/\sqrt{2}$ and the wavelength of this mode is 
$\lambda_0 = {2\pi / k_0}$. The ratio $L/\lambda_0$ gives 
an estimate of the number $N_s$ of TBSs which 
are generated: $N_s = |c| L /(\sqrt{2}\pi)$, 
where $|c|$ is now a function of $n|g|=2\gamma/L$ 
with $\gamma =N|a_s|/a_H$. 
Note that the criterion of modulational instability we use 
has been recently re-derived with a sophisticated 
time-dependent analysis [15]. 
\par
In Figure 4 we plot $N_s/L$ as a function of 
$\gamma/L$ for different values of the anisotropy parameter $\lambda$. 
For small values of $\gamma/L$ we are in 
the weak-coupling quasi-1D regime but approaching $\gamma/L =1/2$ 
our procedure gives a divergent number 
of solitons because one of the two transverse widths shrinks 
to zero for a large attractive inter-atomic interaction. 
Figure 4 shows that the number $N_s$ of TBSs generated 
via modulational instability increases with the anisotropy 
parameter $\lambda$. Moreover, as expected, we find 
$N_s(\lambda)=N_s(\lambda^{-1})$. 
\par 
In conclusion, 
we have analyzed metastable states (triaxial bright solitons) 
and collective oscillations of an attractive Bose 
condensate under anisotropic harmonic confinement. 
We have predicted for the triaxial bright soliton 
a transition from a cigar-like shape 
to a disk-like shape by increasing the inter-atomic strength, 
namely the number of particles or the (negative) scattering length, 
for instance by using Feshbach resonances. 
We have also shown 
that the number $N_s$ of triaxial bright solitons generated 
with a sudden change of the scattering length from positive 
to negative grows with the anisotropy of the 
transverse confinement. Moreover $N_s$ diverges 
at a critical inter-atomic strength, where one of the two 
transverse widths of the soliton train shrinks to zero.

\newpage

\begin{figure}
\centerline{\psfig{file=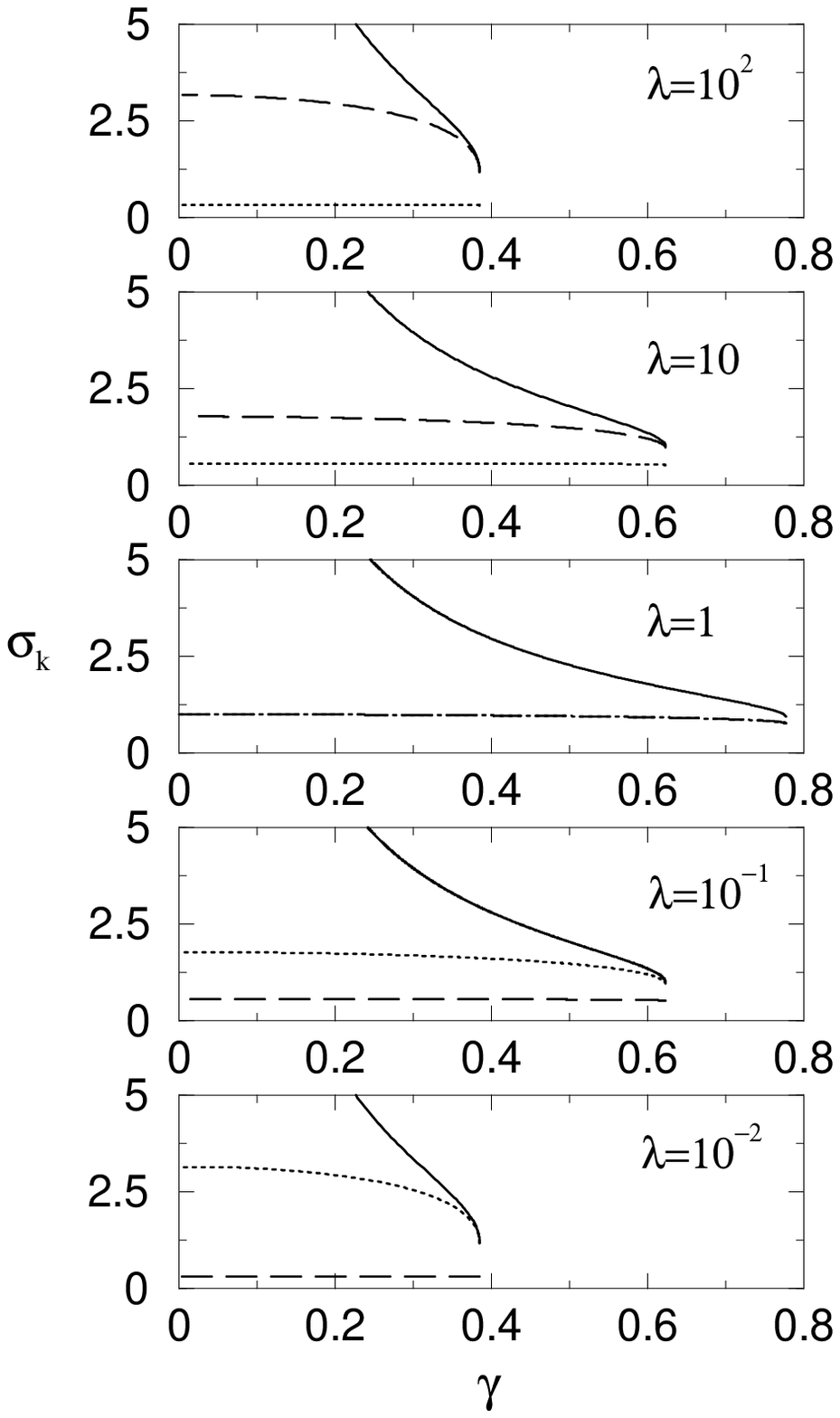,height=6.in}}
\small{FIG. 1: Widths $\sigma_k$ of the TBS as a 
function of the inter-atomic strength $\gamma = N|a_s|/a_H$, 
where $\lambda=\omega_2/\omega_1=\lambda_2/\lambda_1$ 
and $a_H=\left(\hbar/(m\omega_H)\right)^{1/2}$ with 
$\omega_H=(\omega_1\omega_2)^{1/2}$. 
Dotted line: $\sigma_1$, 
dashed line: $\sigma_2$, solid line: $\sigma_3$. 
Lengths are in units of the harmonic length $a_H$ of the 
external transverse potential.} 
\end{figure} 

\newpage 

\begin{figure}
\centerline{\psfig{file=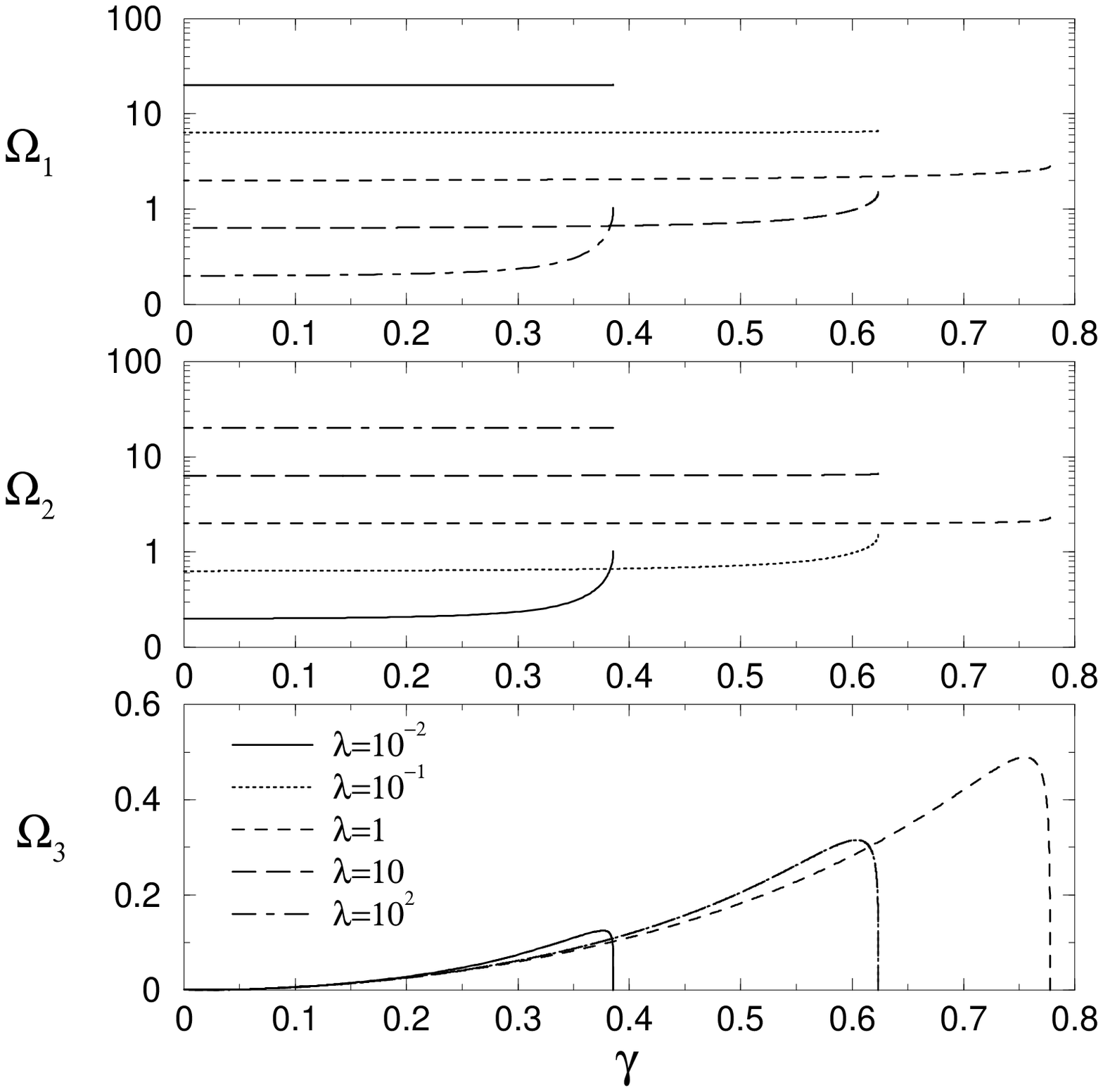,height=4.3in}}
\small{FIG. 2: Collective frequencies $\Omega_k$ of the TBS 
as a function of the inter-atomic strength $\gamma = N|a_s|/a_H$ 
for some values of the anisotropy parameter 
$\lambda=\omega_2/\omega_1=\lambda_2/\lambda_1$. 
Frequencies are in units of the 
harmonic frequency $\omega_H=(\omega_1\omega_2)^{1/2}$ 
of the external transverse potential. } 
\end{figure} 

\newpage 

\begin{figure}
\centerline{\psfig{file=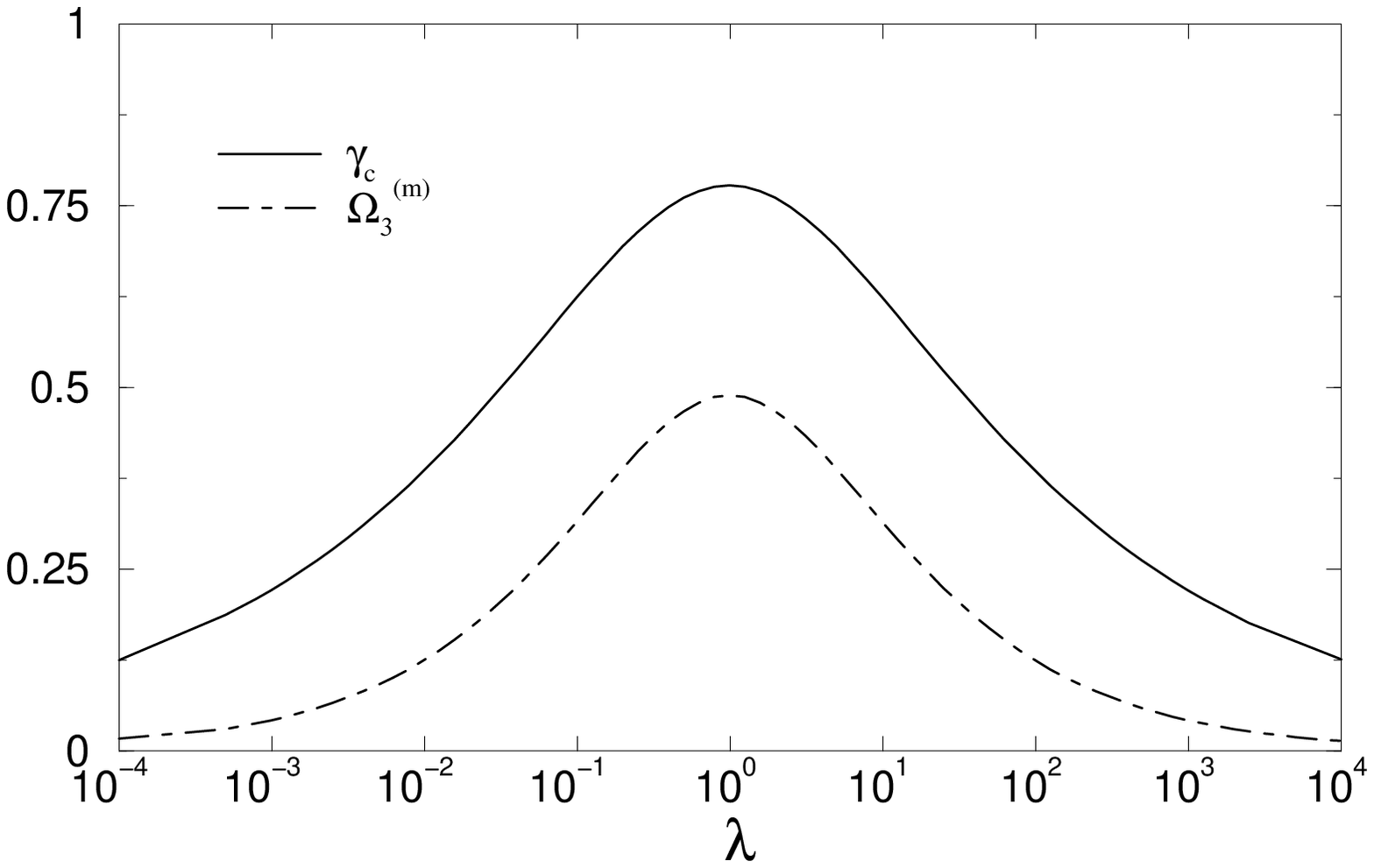,height=3.in}}
\small{FIG. 3: Properties of the TBS at the collapse as a function 
of the anisotropy parameter 
$\lambda=\omega_2/\omega_1=\lambda_2/\lambda_1$. 
The solid line is the critical inter-atomic strength $\gamma_c$. 
The dot-dashed line is the maximum value $\Omega_3^{(m)}$ 
of the collective frequency $\Omega_3$ along the longitudinal $z$ axis. 
Units as in Fig. 2. } 
\end{figure} 

\newpage 

\begin{figure}
\centerline{\psfig{file=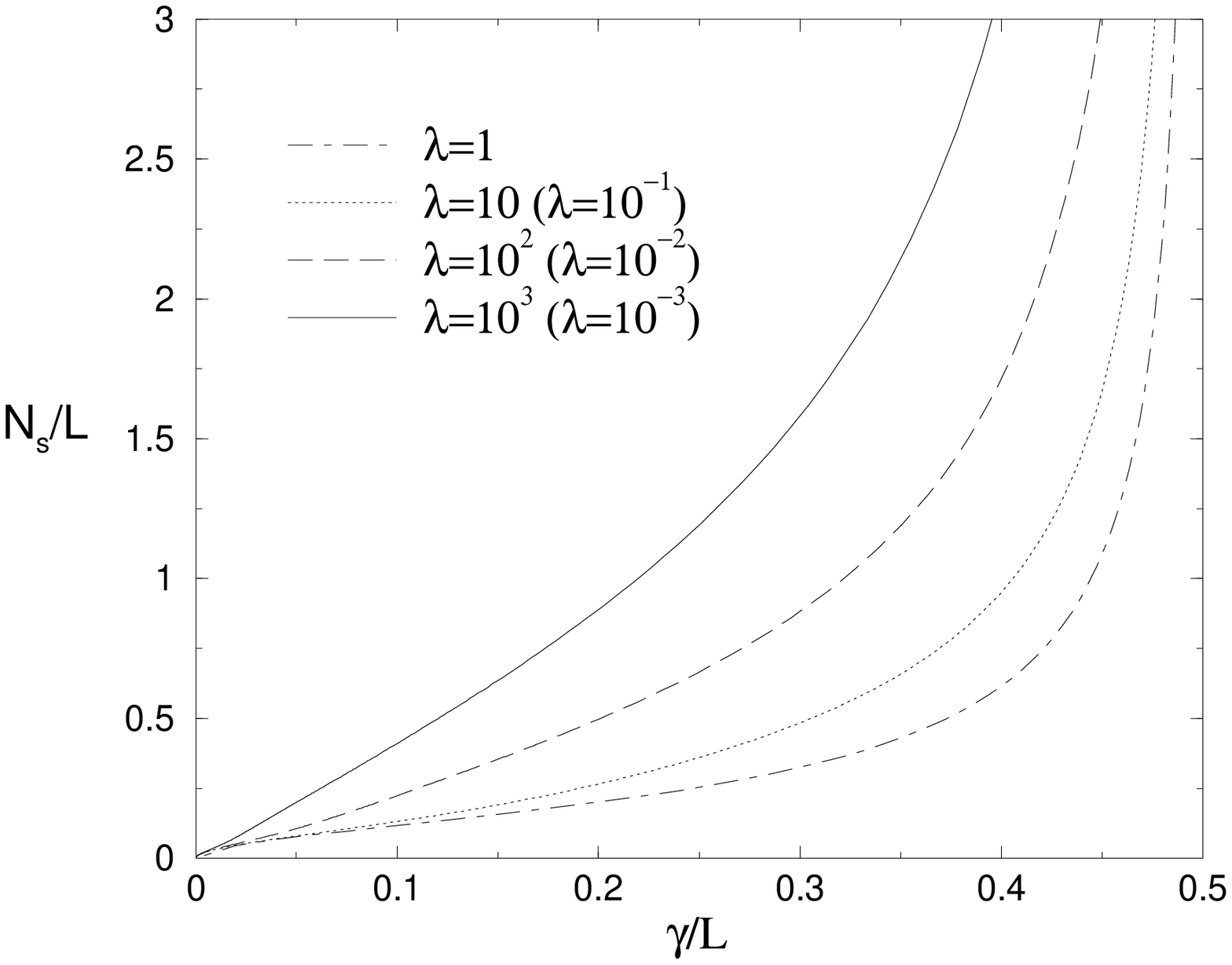,height=3.5in}}
\small{FIG. 4: Number $N_s$ of bright solitons 
generated via modulational instability as a function 
of the inter-atomic strength $\gamma = N|a_s|/a_H$. 
$L$ is the initial axial length of the BEC 
with $N$ atoms. $\lambda=\omega_2/\omega_1=\lambda_2/\lambda_1$ 
is the anisotropy parameter of the transverse harmonic 
confinement. Units as in Fig. 1. } 
\end{figure} 

\end{document}